\documentclass[aps,prl,twocolumn,superscriptaddress]{revtex4}
\usepackage{mathrsfs}
\usepackage{epsfig}
\usepackage{graphicx}
\usepackage{amsfonts}
\usepackage[figuresright]{rotating}
\usepackage{amssymb}
\usepackage{amsmath}
\usepackage{dcolumn}
\usepackage{bm}
\usepackage{color}
\usepackage{xcolor}

\usepackage[pdftex,colorlinks=true]{hyperref}

\def\be{\begin{equation}} \def\ee{\end{equation}}
\def\bea{\begin{eqnarray}} \def\eea{\end{eqnarray}}

\newcommand{\WQCASQC} { Wilczek Quantum Center and Key Laboratory of Artificial Structures and Quantum Control, School of Physics and Astronomy, Shanghai Jiao Tong University, Shanghai 200240, China}

\newcommand{\pku}{State Key Laboratory of Artificial Microstructure and Mesoscopic
Physics, School of Physics, Peking University, 100871 Beijing, China}

\begin{document}
\title{Emergent long-tail dynamics in driven magnets with dynamical  frustration}
\author{Chenyue Guo}
\affiliation{\pku}

\author{Hongzheng Zhao}
\email{hzhao@pku.edu.cn}
\affiliation{\pku}

\author{Zi Cai}
\email{zcai@sjtu.edu.cn}
\affiliation{\WQCASQC}

\begin{abstract}
In this study, we show that dynamical frustration can spontaneously emerge in frustration-free magnetic systems under periodic driving. Specifically, we consider a classical spin system and demonstrate the emergence of spin-ice physics when drive-induced heating is well suppressed. In particular, we focus on the dynamics of magnetic monopole excitations, which, in sharp contrast to their equilibrium counterparts, exhibit a non-ergodic stochastic random-walk process with long-tailed, power-law–distributed waiting times, where the power-law exponent is tunable by the system’s effective temperature. Heating is accelerated at intermediate driving frequencies, and the system eventually heats up to an infinite-temperature state. However, the heating time is extremely sensitive to different initial-state realizations and also follows a long-tailed power-law distribution. We show that a drive-induced short-range attractive interaction between monopoles is responsible for the long-tailed distributions observed in both monopole and heating dynamics.

\end{abstract}


\maketitle

{\it Introduction --} Frustrated magnetism\cite{Misguich2005}  hosts intriguing phenomena ranging from spin glass\cite{Mezard1986}, spin ice\cite{Castelnovo2012} to spin liquid\cite{Lacroix2011}. Typically, frustration arises when  competing interactions in a magnetic system cannot be simultaneously minimized. The resulting  macroscopic degeneracy in the classical groundstate manifold is responsible for a plethora of phenomena such as the order by disorder\cite{Villain1979,Henley1987,Chubukov1992}, fractionalization\cite{Wan2015,Kourtis2016}, and emergent gauge field\cite{Huse2003,Henley2010}. Despite intensive effort, most studies in this regard focus on equilibrium or near equilibrium physics tied to the macroscopic groundstate degeneracy, while the role of frustration remains elusive when a magnetic system is driven far from equilibrium, where traditional (free) energy minimization paradigm is no longer applied. Instead, the interplay between driving and frustration gives rise to richer phenomena\cite{Wan2017,Wan2018,Bittner2020,Jin2022,Yue2023,Hanai2024,Pizzi2025} than what is expected on the basis of these effects separately.

In past decades, periodic (Floquet) driving has not only enriched the toolboxes to manipulate quantum matter~\cite{Tome1990,Sides1998,Oka2009, Lindner2011,Struck2011,Jotzu2014,Zhou2023}, but is also crucial for the appearance of a wealth of phenomena without equilibrium counterpart\cite{khemani2016phase,else2016floquet,yao2017discrete,Martin2017,Huang-Liu:18,Chen2025,Fu2024}. Instead of studying frustrated systems under periodic driving, here we demonstrate that frustration itself could be an emergent property tied to the nonequilibrium feature of periodic driving. By staggering competing interactions, we implement a time-dependent Hamiltonian that is frustration free at any given time. Therefore, frustration plays no role if the switching between the competing interactions is sufficiently slow, where the system keeps evolving under the frustration-free instantaneous Hamiltonian. In contrast, when this periodic switching is too fast to be followed by the system, it will experience a time-averaged Hamiltonian, where frustration dynamically emerges.  

\begin{figure}[htbp]
	\centering
	\includegraphics[width=0.49\textwidth]{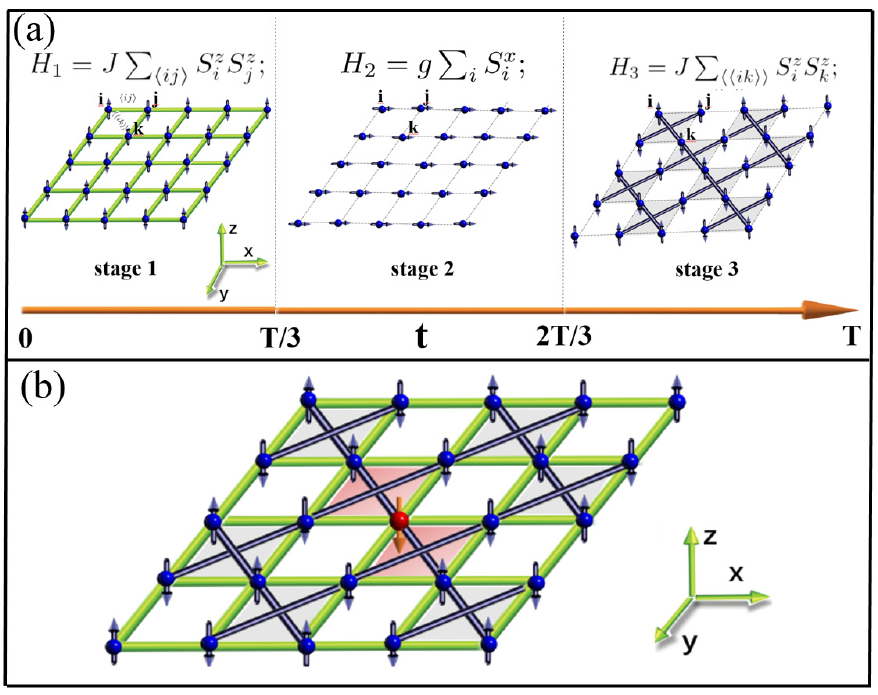}
	\caption{(a)Sketch of the ternary driving protocol within one period and the instantaneous ground state of $H(t)$. The green(blue) solid bonds indicate the AF coupling between the NN (NNN) spins. (b)Sketch of the effective Floquet Hamiltonian in fast driven case. The initial state is prepared as a SI state with a spin flip (red arrow), which generates a pair of monopoles in adjacent plaquettes (pink $\boxtimes$).}
	\label{fig:fig1}
\end{figure}

 A natural question thus arises: how this emergent dynamical frustration differs from its equilibrium counterpart?  To address this question, for concreteness, we consider a Floquet protocol as shown in Fig.~\ref{fig:fig1} (a). A generic Floquet many-body system generally heats up towards an
 infinite temperature state that is entirely
featureless. Yet, with fast driving, a long-lived transient prethermal regime appears where system's energy remains approximately conserved~\cite{ExponentiallyDima,Rigorous2016Mori}. We show that a prototypical frustrated system, spin ice (SI), spontaneously emerges with elementary deconfined monopole-like excitations at low energies~\cite{Castelnovo2008}. 
Our key finding is that, these monopole excitations exhibits a non-ergodic stochastic random walk, characterized by a long-tailed, power-law-distributed waiting time with a divergent average, where the power-law exponent is tunable by the background effective temperature $T_{\mathrm{eff}}$.
This is sharply distinct from the monopole dynamics in the undriven SI, where a standard random walk is generally expected~\cite{Castelnovo2008}. We attribute this anomalous behavior to the drive-induced short-range interaction, which introduces an energy barrier that traps two monopoles upon collision, thereby leading to an extended waiting time before their free motion resumes. 

Heating becomes noticeable at intermediate driving frequencies. Rather surprisingly, the corresponding prethermal lifetimes are extremely sensitive to different initial-state realizations and follow a long-tailed probability distribution. This stands in striking contrast to the conventional understanding of Floquet heating, which is determined by the driving protocol and the initial-state temperature~\cite{Takashi2022Heating}, and thus depends only weakly on the microscopic details of the system.

{\it Model --} The proposed system is a periodically driven classical spin model in a two-dimensional (2D) $L\times L$ lattice.  The Hamiltonian $H(t)$ within one period $T$ reads:
\begin{equation}
	H(t)=\begin{cases}
			J\sum_{\langle ij\rangle}  S_i^z S_j^z;  &  nT <t\leq (n+\frac 13) T\\
			g\sum_i S_i^x;  & ( n+\frac 13) T <t\leq  ( n+\frac 23) T\\
		    J\sum_{\langle\langle ik\rangle\rangle}  S_i^z S_k^z;  &  (n+\frac 23) T <t\leq (n+1) T
		\end{cases} \label{eq:Ham}
	\end{equation}
where $n$ is the number of driving circles and the  spin $\mathbf{S}_i=[S_i^x,S_i^y,S_i^z]$ is a classical unit vector ($|\mathbf{S}_i|=1$) on site i.  $\langle\cdot \cdot\rangle$ indicates the nearest neighboring (NN) bonds in the square lattice, while $\langle\langle \cdot \cdot\rangle\rangle$ represents the diagonal bonds in the grey $\boxtimes$ in Fig.\ref{fig:fig1} (a).  $J>0$ is the strength of antiferromagnetic(AF) coupling and $g$ is the strength of a transverse field along x-direction. In the following, we keep $g\ll J$. Such a small transverse field is sufficient to result in non-trivial dynamics other than the spin procession along z-direction for this interacting spin system.  

The equation of motion for spin $\mathbf{S}_i$  reads:
\begin{equation}
\dot{\mathbf{S}}_i=	 \mathbf{h}_i(t) \times \mathbf{S}_i \label{eq:EOM}
\end{equation}
which describes a spin procession around an effective magnetic field $\mathbf{h}_i(t)$ originating from either the interactions from the neighbors of site i or external transverse field. We note that during each stage of the protocol, $\mathbf{h}_i(t)$ is a constant field along either z or x direction, and hence the corresponding evolution can be integrated analytically, enabling highly efficient large-scale and long-time numerical simulations~\cite{AsymptoticMarin2019,Supplementary}.  

As for experimental realizations of our model,  although it is impossible to switch on and off the interactions between spins in a solid state magnet, recent developments in the Rydberg atomic arrays shed light on this aim. In such  a system,  each Rydberg atom is trapped by an optical tweezer that can be individually moved in a programmable way\cite{Ebadi2022,Bluvstein2023}, which enable us to dynamically modulate the interactions between the Rydberg atoms by tuning their relative distances.  The transverse field can be realized by implementing an laser that coherently excites the atoms from it ground state to  excited state. Despite the similarity, a crucial difference between the Rydberg array and our model is that the former is quantum while the latter is  classical. The effect of quantum fluctuation  will be discussed latter on.

{\it Fast driving: emergence of frustration -}  Next we will study the dynamics with various driving periods.  We first consider the fast driven case with a driving period much smaller than the intrinsic time scales of the system. The stroboscopic dynamics of such a system can be described by a time-independent Floquet Hamiltonian that can be expressed in terms of the Magus expansion as:
\begin{equation}
H_F=H^{(0)}+\frac 1\omega H^{(1)} +\frac 1{\omega^2} H^{(2)}+\dots \label{eq:Floquet}
\end{equation}
where $\omega=\frac{2\pi}{T}$ is the driving frequency. The zero-order approximation $H^{(0)}$ is a time-averaged Hamiltonian over a period: $H^{(0)}=\frac 1T \int_0^T dt H(t)$, which takes the form: 
\begin{equation}
3 H^{(0)}=J \sum_{\boxtimes}\sum_{ij\in \boxtimes } S_i^z S_j^z+ g  \sum_i S_i^x, \label{eq:Ham2}
\end{equation}
where $\boxtimes$ indicates the grey plaquettes in Fig.\ref{fig:fig1} (b).  
$H^{(0)}$ describes a 2D counterpart of SI model. Hence, although the instantaneous Hamiltonian at any given time is frustration-free, our system indeed exhibits an emergent frustration phenomenon at stroboscopic times. At low energies and for small $g$, this effective description features a Coulomb phase with monopole-like excitations\cite{Castelnovo2008}. In particular, the ground states of Hamiltonian.(\ref{eq:Ham2}) with $g=0$ satisfy the local constraint (ice rule): each $\boxtimes$ contains two spins up and two down, leading to $\sum_{i\in\boxtimes} S_i^z=0$.   Flipping a spin in a ground state breaks the ice rule in adjacent  $\boxtimes$ thus creates two monopoles, which can be separated in distance without costing extra energy (deconfined excitations). Therefore, one expects a 2D random walk for the monopoles in the presence of a weak noise, although the monopole dynamics in real spin-ice materials could be more complex\cite{Hallen2022}. 

The relaxation\cite{Jaubert2009,Castelnovo2010} and transport\cite{Bramwell2009,Mostame2014,Pan2016}  properties of monopoles in the SI systems have been studied. A key difference between our model and previous studies is the absence of dissipation in our model, and the monopole dynamics can not be fully captured by the SI Hamiltonian $H^{(0)}$, but involves  higher order terms in the Floquet Hamiltonian.(\ref{eq:Floquet})\cite{Supplementary}.  Although the driving-induced higher order terms are parametrically suppressed, $\mathcal{O}(\omega^{-1})$, in the high-frequency regime, their effects accumulate  and cannot be neglected at long times. Crucially, as we demonstrate below, the monopole evolution is sharply different from the standard random walk, rather, it exhibits long-tail distribution of waiting times, as a direct consequence of the non-equilibrium drive.

 \begin{figure*}[htb]
	\includegraphics[width=0.99\linewidth]{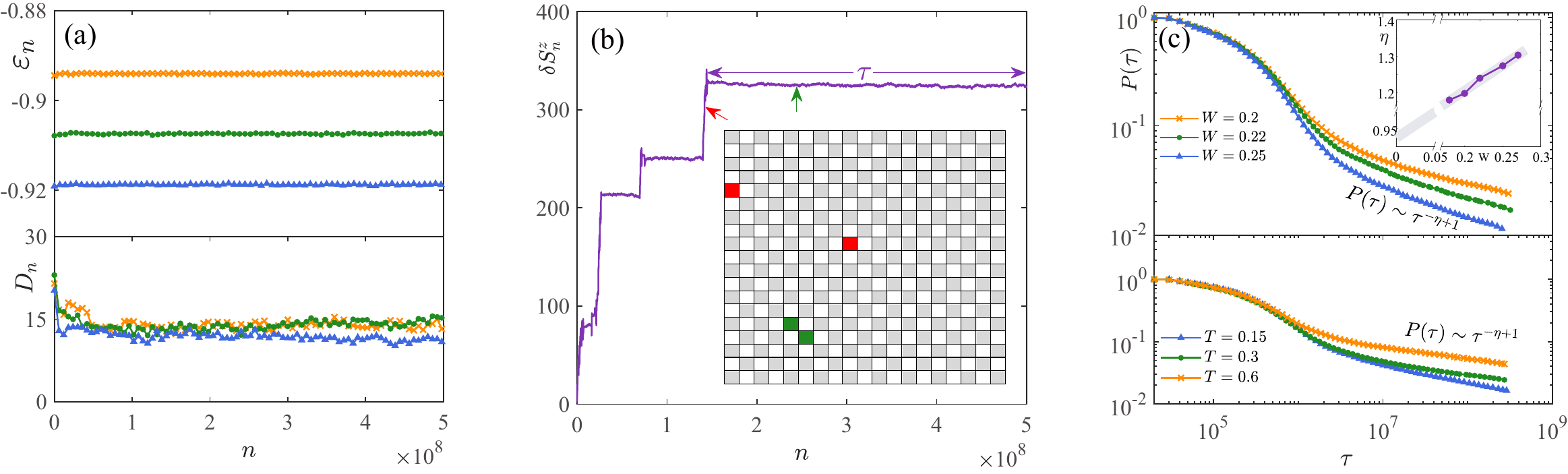}
	\caption{(Color online)(a)Dynamics of the stroboscopic energy per site $\varepsilon_n$ and the spin ice parameter $D_n$ in different trajectories starting from initial states with various noise realizations in fast driven case. (b)The dynamics of the total number of flipped spins $\delta S_z$  during the movement of monopoles in a single trajectory. The inset shows the position of two monopoles at the time slices within a plateau (green) and between two plateaus (red). (c)$P(\tau)$  with a  fixed $T=0.3J^{-1}$ but various $W$ (upper panel) and a fixed $W=0.2J$ but various $T$ (lower panel). The statistics are performed over $\mathcal{N}=500$ trajectories. The inset shows the noise strength $W$ dependence of $\eta$. The driving period is chosen as $T=0.3J^{-1}$ for (a) and (b). The system size is $L=20$.} \label{fig:fig2}
\end{figure*}

To prepare the initial state, we start from a ground state of Hamiltonian.(\ref{eq:Ham2}), then create a pair of monopoles on top of it by randomly flipping a spin. To mobilize these monopoles, we perturb this initial state by imposing a noise $\delta_i^z$ with an uniform random distribution within $[-W ,W]$ on each $S_i^z$\cite{Supplementary}. To perform the statistical average,  we choose $\mathcal{N}$ initial states with different noise realizations. We focus on the stroboscopic dynamics  at the time slices $t_n=n T$ with a fast driving ($T=0.3J^{-1}$). 

Before presenting monopole dynamics, we first consider the stroboscopic energy per site $\varepsilon_n=H(t_n)/L^2$ and the ice rule parameter: $D_n=\sum_{\boxtimes} |\sum_{i\in\boxtimes} S_i^z(t_n)|$, the former is used to measure the energy absorbed from the periodic driving, while the latter can characterize the violation of the ice rule, which is related to the number of monopoles ({\it e.g.} $D=4$ for a perfect SI state with two monopoles).  As shown in Fig.\ref{fig:fig2} (a), starting from different initial states, system's energy $\varepsilon_n$ remains close to its initial value within our simulated time scale. Therefore, the system enters a prethermal regime where the stroboscopic dynamics is well captured by the time-independent Floquet Hamiltonian (\ref{eq:Floquet}), leading to the emergent frustration phenomenon. Consequently, $D_n$ also remains approximately unchanged, 
suggesting that the number of monopoles is approximately conserved. Note that initially, $D_n >4$ due to the presence of the initial state noise in the spin configuration.

 \begin{figure*}[htbp]
	\centering
	\includegraphics[width=0.99\textwidth]{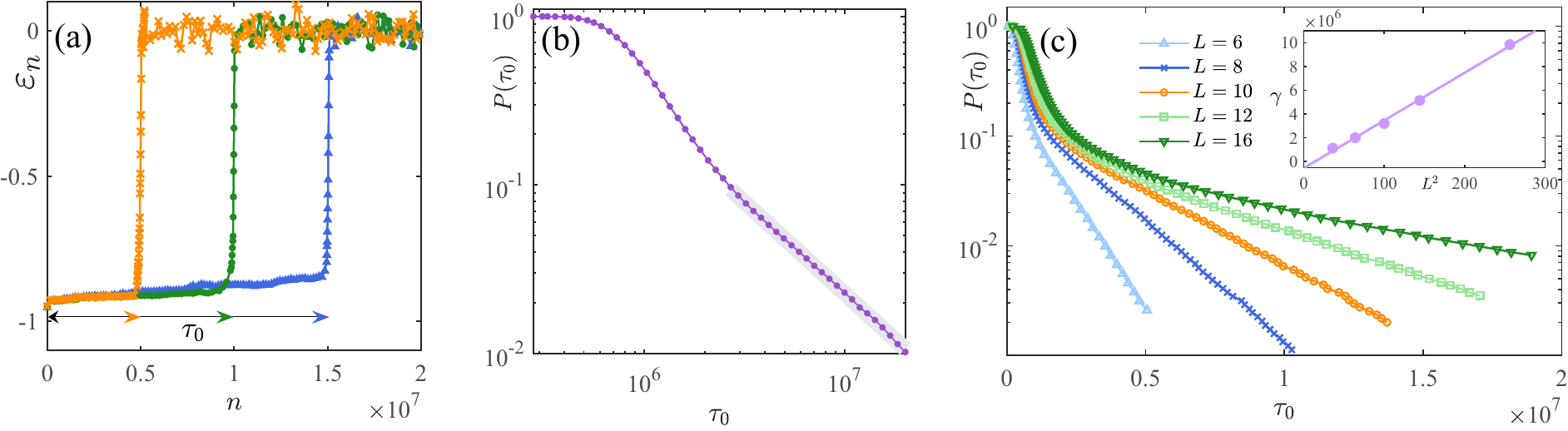}
	\caption{(a)Dynamics of  $\varepsilon_n$ in different trajectories starting from initial states with various noise realizations in the presence of intermediate driving. (b)The cumulative distribution function of the prethermal plateau duration $P(\tau_0)$, which also exhibits a power-law decay for large $\tau_0$ (log-log plot). (c)$P(\tau_0)$ for small systems (semi-log plot), which exhibit an exponential decay for large $\tau_0$: $e^{-\tau_0/\gamma}$. The statistics are performed over $\mathcal{N}=2\times10^4$ trajectories. The inset shows the size dependence of $\gamma$. The parameters are chosen as $L=20$, $T=5.2J^{-1}$, $W=0.1J$ for (a)-(c).}
	\label{fig:fig3}
\end{figure*}

{\it Long-tail monopole dynamics with emergent short-range interactions-} To monitor the monopole trajectories, we define $\delta S_i^z=S_i^z(t_n)-\tilde{S}_i^z(t_n)$ to measure the difference between the two spin states at $t=t_n$ starting from the initial states {\it with} and {\it without} the initial spin flip. Since the movement of a monopole is accompanied by flipping a sequence of spins with alternating $S_i^z$, we further define $\delta S_n^{z}=\sum_i |\delta S_i^z(t_n)|$ as the total number of the flipped spins during the movement to measure the length of the monopole trajectories. In Fig.\ref{fig:fig2} (b), we show a typical dynamical behavior of $\delta S_n^{z}$ in a single trajectory. 
Multiple plateaus with exceptionally long duration appear ({\it e.g.} $n\sim 10^8$), where $\delta S_n^{z}$ does not change in time and hence {\it the monopoles are stuck}, 
while the jump between these plateaus indicates the movement of monopoles. 

Crucially, these long-lived plateaus occur only when {\it two monopoles collide and stick to each other}. To see this, we calculate the SI parameter in each plaquette: $D_n^\boxtimes=|\sum_{i\in\boxtimes} S_i^z(t_n)|$ ($D_n^\boxtimes\approx 2$ indicates a monopole located in plaquette $\boxtimes$) to locate the position of a monopole\cite{Supplementary}.  As shown in the inset of Fig.\ref{fig:fig2} (b), when monoples are mobile, they behave like free particles wandering in the lattice and can be well separated ({\it e.g.} red plaquettes at a time slice between different plateaus). 
In contrast, two monopoles collide and stick to each other during the plateau (green plaquettes) for extraordinarily long (e.g. $n \sim 10^8$). We also note that{ monopoles do not annihilate because there is no dissipation}.  These results imply an emergent attractive interactions between the monopoles with a force range compatible to the lattice constant, beyond which this interaction plays no role.
After the plateau, monopoles can separate and keep wandering again until their next collision.

To reveal quantitatively the consequence of this emergent force, we perform statistical analysis over the duration $\tau$ of these plateaus in $\mathcal{N}=500$ trajectories, and the probability density of the duration of these plateaus is defined as  $p(\tau)$. To reduce the statistical error, we further consider the complementary cumulative distribution function $P(\tau)=\int_{\tau}^\infty p(\tau') d\tau'$ instead of $p(\tau)$, which quantifies the probability of finding a plateau whose duration is larger than $\tau$ ({\it e.g.} $P(0)=1$). 
As shown in  Fig.\ref{fig:fig2} (c), for those long-lived plateaus, $P(\tau)\sim \tau^{-\eta+1} $ is observed suggesting a long-tail power-law distribution $p(\tau)\sim \tau^{-\eta}$ with a non-universal power $\eta>1$.

This long-tail distribution arises due to the energy barrier 
$V$ induced by their attractive force, that the two monopoles must overcome before moving freely. The typical trapping time $\tau$ satisfies  the  Arrhenius law: 
$\tau\sim e^{V/T_{\mathrm{eff}}}$, where $T_{\mathrm{eff}}$ is an effective temperature. We also note that $V$ in our model is not a constant but a random number. This is because a monopole in our setting is an emergent quasiparticle— a local excitation on top of the SI background with weak fluctuations. The underlying spin configuration can vary in both space and time, and consequently, the shape and size of the monopole, as well as its interaction with the SI background, also change. This effectively generates energy barrier that fluctuates as well, and for simplicity here we assume that it follows the Poisson distribution: $p(V)\sim e^{-\alpha V}$.
This naturally leads to the probability density of $\tau$:
\begin{equation}
\label{eq.T-dependence}
p(\tau)=\int dV~\delta(\tau-e^{\frac{V}{ T_{\mathrm {eff}}}}) p(V)\sim \tau^{-1-\alpha  T_{\mathrm{eff}}},
\end{equation} 
thus one identifies that $\eta=1+\alpha  T_{\mathrm {eff}}$. As detailed in \cite{Supplementary}, in our system $T_{\mathrm{eff}}$ is proportional to the initial noise strength $W$. Therefore, Eq.~\eqref{eq.T-dependence} is in accordance with our numerical observation shown in the inset of Fig.\ref{fig:fig2} (c), where we find $\eta=a+b W$ with $a=0.95\pm 0.1$ and $\eta<2$ for typical fast driving period. A power-law distribution  $p(\tau)\sim \tau^{-\eta}$ with $1<\eta<2$ indicates a divergence of the average trapping time $\langle\tau\rangle$, which resembles the non-ergodic anomalous diffusion in the continuous time random walk with long-tail distributed waiting time\cite{Igor2008}. Also, it reminds us of self-organized critical systems, where the probability of large events (e.g., earthquakes or avalanches) decays algebraically with their intensity\cite{Bak1987}.



 
The emergent short-range attractive interaction can be attributed to the drive-induced higher-order terms, which can be perturbatively obtained using the Magnus expansion of the Floquet Hamiltonian.(\ref{eq:Floquet}). They involve {\it local} multi-spin interactions\cite{Supplementary} thus give rise to effective {\it short-range} interactions between the monopoles. To verify the relation between this emergent interaction and the high-order terms in Hamiltonian.(\ref{eq:Floquet}), we compare $P(\tau)$ with different driving period $T$. As shown in the lower panel of Fig.\ref{fig:fig2} (c), $P(\tau)$ decay faster for smaller $T$, indicating shorter $\tau$ on average,  which, on the other hand, implies weaker attractive interactions $V$.  This result agrees with the Magnus expansion in Eq.(\ref{eq:Floquet}) where the contribution of higher order terms is suppressed for small $T$. 

{\it Intermediate driving: heating dynamics with long-tail distribution --}  
Next, we focus on an intermediate frequency regime  ($T{\sim} J^{-1}$) where heating is accelerated. As shown in Fig.\ref{fig:fig3} (a), for different trajectories starting from initial states with various noise realizations, the stroboscopic energy $\varepsilon_n$ experiences a  plateau with duration $\tau_0$, after which it quickly increases to its infinite temperature value. $\tau_0$ significantly varies from one trajectory to another, and could be extraordinarily long ($n{\sim} 10^7$) in some trajectories.  In Fig.\ref{fig:fig3} (b), we plot $P(\tau_0){=}\int_{\tau_0}^\infty p(\tau'_0) d\tau'_0$, where $p(\tau_0)$ is the probability density of $\tau_0$, which also decays algebraically with $\tau_0$ for large $\tau_0$, indicating again, a long-tail distribution. This thus suggests a strong initial-state dependence of the heating time, in contrast to generic Floquet many-body systems, where heating occurs on a typical time scale that depends only on the driving protocol and initial-state
temperature~\cite{Rigorous2016Mori}.


Crucial insights about the heating dynamics can be gained from the finite-size effect of $P(\tau_0)$. To this end, we calculate $P(\tau_0)$ for smaller systems, where we can find that the algebraic decay of $P(\tau_0)$ for large $\tau_0$ is replaced by an exponential decay $P(\tau_0){\sim} e^{-\tau_0/\gamma}$ (see Fig.\ref{fig:fig3} c), where $\gamma$ is a characteristic time scale of heating that strongly depends on the system size $L$. As shown in the inset of Fig.\ref{fig:fig3} (c), the finite-size scaling of $\gamma$ suggests $\gamma$  diverges with $L$ as  $\gamma{\sim} L^2$, and when $L{\to}\infty$ the exponential decay induced by the finite-size effect gives ways to a power-law decay.

 These numerical results suggest that the physical picture of monopole wandering and trapping also applies here. As detailed in Ref.~\cite{Supplementary}, we find that heating is typically initiated by the collision of two monopoles. Instead of being perfectly localized quasiparticles as in the fast-driven case, the two monopoles here smear out and remain bound together for a long time. They generate a region of high energy density, thereby opening new heating channels that enable rapid energy absorption. Hence, heating spreads from a “seed” (the location of the collision) to the entire lattice, and the ice rule is also broken accordingly. Before heating begins, the two monopoles wander through the lattice, and the typical timescale for their collision during this 2D random walk scales as $L^2$, which explains the finite-size dependence of $P(\tau_0)$.

{\it Conclusion and outlook --} In summary, we propose a dynamical frustration mechanism in periodically driven systems, which exhibit a long-tailed, power-law distribution of the waiting time for monopole random walks. Drive-induced processes play a key role in monopole dynamics under fast and intermediate driving: they lead to a short-range attraction between monopoles and trap them in the former, while initiating heating at late times in the latter. These emergent properties appear only when heating is well suppressed, and no dynamical frustration is observed under slow driving (see details in \cite{Supplementary}). 

We focus only on the dynamics of a pair of monopoles, while the situation with a finite density of monopoles may be even more interesting, as the emergent attractive interactions could lead to unprecedented collective behaviors absent in conventional spin-ice systems. 

Although the proposed driving protocol might be difficult to realize in solid-state systems, it is already accessible using Rydberg quantum simulators~\cite{Ebadi2022,Bluvstein2023,homeier2025prethermal}.  We expect that similar attractive interactions will also appear in quantum systems, and a natural
question is whether the long-tail dynamics still persists. In that case, quantum tunneling will make it easier for the monopoles to escape from the attractive potential, thereby reducing their trapping time. The competition between drive-induced attraction and quantum tunneling may significantly enrich the possible forms of anomalous monopole dynamics. Moreover, quantum interference between different monopole paths might qualitatively alter their long-time dynamics and lead to anomalous diffusion.

{\it Acknowledgments}.--- 
We thank Roderich Moessner and Andrea Pizzi for stimulating discussions.
ZC is supported by the National Key Research and Development Program of China (2024YFA1408303,2020YFA0309000), Natural Science Foundation of China (Grant No.12174251 and No.12525407),  Shanghai Municipal Science and Technology Major Project (Grant No.2019SHZDZX01), Shanghai Science and Technology Innovation Action Plan(Grant No. 24Z510205936). CYG is supported by the the China Postdoctoral Science Foundation (Grant No. GZC20252199). 
HZ is supported by Quantum Science and Technology-National Science and Technology Major Project
(No. 2024ZD0301800) and by the National Natural Science
Foundation of China (Grant No. 12474214), and by ``High-performance Computing Platform of Peking University".


\begin{thebibliography}{49}
	\expandafter\ifx\csname natexlab\endcsname\relax\def\natexlab#1{#1}\fi
	\expandafter\ifx\csname bibnamefont\endcsname\relax
	\def\bibnamefont#1{#1}\fi
	\expandafter\ifx\csname bibfnamefont\endcsname\relax
	\def\bibfnamefont#1{#1}\fi
	\expandafter\ifx\csname citenamefont\endcsname\relax
	\def\citenamefont#1{#1}\fi
	\expandafter\ifx\csname url\endcsname\relax
	\def\url#1{\texttt{#1}}\fi
	\expandafter\ifx\csname urlprefix\endcsname\relax\def\urlprefix{URL }\fi
	\providecommand{\bibinfo}[2]{#2}
	\providecommand{\eprint}[2][]{\url{#2}}
	
	\bibitem[{\citenamefont{G.Misguich and C.Lhuillier}(2005)}]{Misguich2005}
	\bibinfo{author}{\bibnamefont{G.Misguich}} \bibnamefont{and}
	\bibinfo{author}{\bibnamefont{C.Lhuillier}}, \emph{\bibinfo{title}{Frustrated
			spin systems}} (\bibinfo{publisher}{~World Scientific, Singapore},
	\bibinfo{year}{2005}).
	
	\bibitem[{\citenamefont{Mezard et~al.}(1986)\citenamefont{Mezard, Parisi, and
			Virasoro}}]{Mezard1986}
	\bibinfo{author}{\bibfnamefont{M.}~\bibnamefont{Mezard}},
	\bibinfo{author}{\bibfnamefont{G.}~\bibnamefont{Parisi}}, \bibnamefont{and}
	\bibinfo{author}{\bibfnamefont{M.}~\bibnamefont{Virasoro}},
	\emph{\bibinfo{title}{Spin Glass Theory and Beyond: An Introduction to the
			Replica Method and Its Applications}} (\bibinfo{publisher}{~World
		Scientific}, \bibinfo{year}{1986}).
	
	\bibitem[{\citenamefont{C.Castelnovo et~al.}(2012)\citenamefont{C.Castelnovo,
			R.Moessner, and Sondhi}}]{Castelnovo2012}
	\bibinfo{author}{\bibnamefont{C.Castelnovo}},
	\bibinfo{author}{\bibnamefont{R.Moessner}}, \bibnamefont{and}
	\bibinfo{author}{\bibfnamefont{S.}~\bibnamefont{Sondhi}},
	\bibinfo{journal}{Annu. Rev. Condens. Mat. Phys.}
	\textbf{\bibinfo{volume}{3}}, \bibinfo{pages}{35} (\bibinfo{year}{2012}).
	
	\bibitem[{\citenamefont{Lacroix et~al.}(2011)\citenamefont{Lacroix, Mendels,
			and Mila}}]{Lacroix2011}
	\bibinfo{author}{\bibfnamefont{C.}~\bibnamefont{Lacroix}},
	\bibinfo{author}{\bibfnamefont{P.}~\bibnamefont{Mendels}}, \bibnamefont{and}
	\bibinfo{author}{\bibfnamefont{F.}~\bibnamefont{Mila}},
	\emph{\bibinfo{title}{Introduction to Frustrated Magnetism: Materials,
			Experiments, Theory}} (\bibinfo{publisher}{Springer}, \bibinfo{year}{2011}).
	
	\bibitem[{\citenamefont{Villain}(1979)}]{Villain1979}
	\bibinfo{author}{\bibfnamefont{J.}~\bibnamefont{Villain}}, \bibinfo{journal}{Z.
		Phys. B} \textbf{\bibinfo{volume}{33}}, \bibinfo{pages}{31}
	(\bibinfo{year}{1979}).
	
	\bibitem[{\citenamefont{Henley}(1997)}]{Henley1987}
	\bibinfo{author}{\bibfnamefont{C.~L.} \bibnamefont{Henley}},
	\bibinfo{journal}{J. Appl. Phys.} \textbf{\bibinfo{volume}{61}},
	\bibinfo{pages}{3962} (\bibinfo{year}{1997}).
	
	\bibitem[{\citenamefont{Chubukov}(1992)}]{Chubukov1992}
	\bibinfo{author}{\bibfnamefont{A.}~\bibnamefont{Chubukov}},
	\bibinfo{journal}{Phys. Rev. Lett.} \textbf{\bibinfo{volume}{69}},
	\bibinfo{pages}{832} (\bibinfo{year}{1992}).
	
	\bibitem[{\citenamefont{Wan et~al.}(2016)\citenamefont{Wan, Carrasquilla, and
			Melko}}]{Wan2015}
	\bibinfo{author}{\bibfnamefont{Y.}~\bibnamefont{Wan}},
	\bibinfo{author}{\bibfnamefont{J.}~\bibnamefont{Carrasquilla}},
	\bibnamefont{and} \bibinfo{author}{\bibfnamefont{R.~G.} \bibnamefont{Melko}},
	\bibinfo{journal}{Phys. Rev. Lett.} \textbf{\bibinfo{volume}{116}},
	\bibinfo{pages}{167202} (\bibinfo{year}{2016}).
	
	\bibitem[{\citenamefont{Kourtis and Castelnovo}(2016)}]{Kourtis2016}
	\bibinfo{author}{\bibfnamefont{S.}~\bibnamefont{Kourtis}} \bibnamefont{and}
	\bibinfo{author}{\bibfnamefont{C.}~\bibnamefont{Castelnovo}},
	\bibinfo{journal}{Phys. Rev. B} \textbf{\bibinfo{volume}{94}},
	\bibinfo{pages}{104401} (\bibinfo{year}{2016}).
	
	\bibitem[{\citenamefont{Huse et~al.}(2003)\citenamefont{Huse, Krauth, Moessner,
			and Sondhi}}]{Huse2003}
	\bibinfo{author}{\bibfnamefont{D.~A.} \bibnamefont{Huse}},
	\bibinfo{author}{\bibfnamefont{W.}~\bibnamefont{Krauth}},
	\bibinfo{author}{\bibfnamefont{R.}~\bibnamefont{Moessner}}, \bibnamefont{and}
	\bibinfo{author}{\bibfnamefont{S.~L.} \bibnamefont{Sondhi}},
	\bibinfo{journal}{Phys. Rev. Lett.} \textbf{\bibinfo{volume}{91}},
	\bibinfo{pages}{167004} (\bibinfo{year}{2003}).
	
	\bibitem[{\citenamefont{Henley}(2010)}]{Henley2010}
	\bibinfo{author}{\bibfnamefont{C.}~\bibnamefont{Henley}},
	\bibinfo{journal}{Annu. Rev. Condens. Mat. Phys.}
	\textbf{\bibinfo{volume}{1}}, \bibinfo{pages}{179} (\bibinfo{year}{2010}).
	
	\bibitem[{\citenamefont{Wan and Moessner}(2017)}]{Wan2017}
	\bibinfo{author}{\bibfnamefont{Y.}~\bibnamefont{Wan}} \bibnamefont{and}
	\bibinfo{author}{\bibfnamefont{R.}~\bibnamefont{Moessner}},
	\bibinfo{journal}{Phys. Rev. Lett.} \textbf{\bibinfo{volume}{119}},
	\bibinfo{pages}{167203} (\bibinfo{year}{2017}).
	
	\bibitem[{\citenamefont{Wan and Moessner}(2018)}]{Wan2018}
	\bibinfo{author}{\bibfnamefont{Y.}~\bibnamefont{Wan}} \bibnamefont{and}
	\bibinfo{author}{\bibfnamefont{R.}~\bibnamefont{Moessner}},
	\bibinfo{journal}{Phys. Rev. B} \textbf{\bibinfo{volume}{98}},
	\bibinfo{pages}{184432} (\bibinfo{year}{2018}).
	
	\bibitem[{\citenamefont{Bittner et~al.}(2020)\citenamefont{Bittner,
			Gole\ifmmode~\check{z}\else \v{z}\fi{}, Eckstein, and Werner}}]{Bittner2020}
	\bibinfo{author}{\bibfnamefont{N.}~\bibnamefont{Bittner}},
	\bibinfo{author}{\bibfnamefont{D.}~\bibnamefont{Gole\ifmmode~\check{z}\else
			\v{z}\fi{}}}, \bibinfo{author}{\bibfnamefont{M.}~\bibnamefont{Eckstein}},
	\bibnamefont{and} \bibinfo{author}{\bibfnamefont{P.}~\bibnamefont{Werner}},
	\bibinfo{journal}{Phys. Rev. B} \textbf{\bibinfo{volume}{102}},
	\bibinfo{pages}{235169} (\bibinfo{year}{2020}).
	
	\bibitem[{\citenamefont{Jin et~al.}(2022)\citenamefont{Jin, Pizzi, and
			Knolle}}]{Jin2022}
	\bibinfo{author}{\bibfnamefont{H.-K.} \bibnamefont{Jin}},
	\bibinfo{author}{\bibfnamefont{A.}~\bibnamefont{Pizzi}}, \bibnamefont{and}
	\bibinfo{author}{\bibfnamefont{J.}~\bibnamefont{Knolle}},
	\bibinfo{journal}{Phys. Rev. B} \textbf{\bibinfo{volume}{106}},
	\bibinfo{pages}{144312} (\bibinfo{year}{2022}).
	
	\bibitem[{\citenamefont{Yue and Cai}(2023)}]{Yue2023}
	\bibinfo{author}{\bibfnamefont{M.}~\bibnamefont{Yue}} \bibnamefont{and}
	\bibinfo{author}{\bibfnamefont{Z.}~\bibnamefont{Cai}},
	\bibinfo{journal}{Phys. Rev. Lett.} \textbf{\bibinfo{volume}{131}},
	\bibinfo{pages}{056502} (\bibinfo{year}{2023}).
	
	\bibitem[{\citenamefont{Hanai}(2024)}]{Hanai2024}
	\bibinfo{author}{\bibfnamefont{R.}~\bibnamefont{Hanai}},
	\bibinfo{journal}{Phys. Rev. X} \textbf{\bibinfo{volume}{14}},
	\bibinfo{pages}{011029} (\bibinfo{year}{2024}).
	
	\bibitem[{\citenamefont{{Pizzi} et~al.}(2025)\citenamefont{{Pizzi},
			{Castelnovo}, and {Knolle}}}]{Pizzi2025}
	\bibinfo{author}{\bibfnamefont{A.}~\bibnamefont{{Pizzi}}},
	\bibinfo{author}{\bibfnamefont{C.}~\bibnamefont{{Castelnovo}}},
	\bibnamefont{and} \bibinfo{author}{\bibfnamefont{J.}~\bibnamefont{{Knolle}}},
	\bibinfo{journal}{arXiv e-prints} \bibinfo{eid}{arXiv:2508.04763}
	(\bibinfo{year}{2025}), \eprint{2508.04763}.
	
	\bibitem[{\citenamefont{Tom\'e and de~Oliveira}(1990)}]{Tome1990}
	\bibinfo{author}{\bibfnamefont{T.}~\bibnamefont{Tom\'e}} \bibnamefont{and}
	\bibinfo{author}{\bibfnamefont{M.~J.} \bibnamefont{de~Oliveira}},
	\bibinfo{journal}{Phys. Rev. A} \textbf{\bibinfo{volume}{41}},
	\bibinfo{pages}{4251} (\bibinfo{year}{1990}).
	
	\bibitem[{\citenamefont{Sides et~al.}(1998)\citenamefont{Sides, Rikvold, and
			Novotny}}]{Sides1998}
	\bibinfo{author}{\bibfnamefont{S.~W.} \bibnamefont{Sides}},
	\bibinfo{author}{\bibfnamefont{P.~A.} \bibnamefont{Rikvold}},
	\bibnamefont{and} \bibinfo{author}{\bibfnamefont{M.~A.}
		\bibnamefont{Novotny}}, \bibinfo{journal}{Phys. Rev. Lett.}
	\textbf{\bibinfo{volume}{81}}, \bibinfo{pages}{834} (\bibinfo{year}{1998}).
	
	\bibitem[{\citenamefont{Oka and Aoki}(2009)}]{Oka2009}
	\bibinfo{author}{\bibfnamefont{T.}~\bibnamefont{Oka}} \bibnamefont{and}
	\bibinfo{author}{\bibfnamefont{H.}~\bibnamefont{Aoki}},
	\bibinfo{journal}{Phys. Rev. B} \textbf{\bibinfo{volume}{79}},
	\bibinfo{pages}{081406} (\bibinfo{year}{2009}).
	
	\bibitem[{\citenamefont{Lindner et~al.}(2011)\citenamefont{Lindner, Refael, and
			Galitski}}]{Lindner2011}
	\bibinfo{author}{\bibfnamefont{N.~H.} \bibnamefont{Lindner}},
	\bibinfo{author}{\bibfnamefont{G.}~\bibnamefont{Refael}}, \bibnamefont{and}
	\bibinfo{author}{\bibfnamefont{V.}~\bibnamefont{Galitski}},
	\bibinfo{journal}{Nature Physics} \textbf{\bibinfo{volume}{7}},
	\bibinfo{pages}{490} (\bibinfo{year}{2011}).
	
	\bibitem[{\citenamefont{Struck et~al.}(2011)\citenamefont{Struck, Olschlager,
			Targat, Soltan-Panahi, Eckardt, Lewenstein, Windpassinger, and
			Sengstock}}]{Struck2011}
	\bibinfo{author}{\bibfnamefont{J.}~\bibnamefont{Struck}},
	\bibinfo{author}{\bibfnamefont{C.}~\bibnamefont{Olschlager}},
	\bibinfo{author}{\bibfnamefont{R.~L.} \bibnamefont{Targat}},
	\bibinfo{author}{\bibfnamefont{P.}~\bibnamefont{Soltan-Panahi}},
	\bibinfo{author}{\bibfnamefont{A.}~\bibnamefont{Eckardt}},
	\bibinfo{author}{\bibfnamefont{M.}~\bibnamefont{Lewenstein}},
	\bibinfo{author}{\bibfnamefont{P.}~\bibnamefont{Windpassinger}},
	\bibnamefont{and}
	\bibinfo{author}{\bibfnamefont{K.}~\bibnamefont{Sengstock}},
	\bibinfo{journal}{Science} \textbf{\bibinfo{volume}{333}},
	\bibinfo{pages}{996} (\bibinfo{year}{2011}).
	
	\bibitem[{\citenamefont{Jotzu et~al.}(2014)\citenamefont{Jotzu, Messer,
			Desbuquois, Lebrat, Uehlinger, Greif, and Esslinger}}]{Jotzu2014}
	\bibinfo{author}{\bibfnamefont{G.}~\bibnamefont{Jotzu}},
	\bibinfo{author}{\bibfnamefont{M.}~\bibnamefont{Messer}},
	\bibinfo{author}{\bibfnamefont{R.}~\bibnamefont{Desbuquois}},
	\bibinfo{author}{\bibfnamefont{M.}~\bibnamefont{Lebrat}},
	\bibinfo{author}{\bibfnamefont{T.}~\bibnamefont{Uehlinger}},
	\bibinfo{author}{\bibfnamefont{D.}~\bibnamefont{Greif}}, \bibnamefont{and}
	\bibinfo{author}{\bibfnamefont{T.}~\bibnamefont{Esslinger}},
	\bibinfo{journal}{Nature} \textbf{\bibinfo{volume}{515}},
	\bibinfo{pages}{237} (\bibinfo{year}{2014}).
	
	\bibitem[{\citenamefont{Zhou et~al.}(2023)\citenamefont{Zhou, Bao, Fan, Zhou,
			Gao, Zhong, Lin, Liu, Yu, Tang et~al.}}]{Zhou2023}
	\bibinfo{author}{\bibfnamefont{S.}~\bibnamefont{Zhou}},
	\bibinfo{author}{\bibfnamefont{C.}~\bibnamefont{Bao}},
	\bibinfo{author}{\bibfnamefont{B.}~\bibnamefont{Fan}},
	\bibinfo{author}{\bibfnamefont{H.}~\bibnamefont{Zhou}},
	\bibinfo{author}{\bibfnamefont{Q.}~\bibnamefont{Gao}},
	\bibinfo{author}{\bibfnamefont{H.}~\bibnamefont{Zhong}},
	\bibinfo{author}{\bibfnamefont{T.}~\bibnamefont{Lin}},
	\bibinfo{author}{\bibfnamefont{H.}~\bibnamefont{Liu}},
	\bibinfo{author}{\bibfnamefont{P.}~\bibnamefont{Yu}},
	\bibinfo{author}{\bibfnamefont{P.}~\bibnamefont{Tang}}, \bibnamefont{et~al.},
	\bibinfo{journal}{Nature} \textbf{\bibinfo{volume}{614}}, \bibinfo{pages}{75}
	(\bibinfo{year}{2023}).
	
	\bibitem[{\citenamefont{Khemani et~al.}(2016)\citenamefont{Khemani, Lazarides,
			Moessner, and Sondhi}}]{khemani2016phase}
	\bibinfo{author}{\bibfnamefont{V.}~\bibnamefont{Khemani}},
	\bibinfo{author}{\bibfnamefont{A.}~\bibnamefont{Lazarides}},
	\bibinfo{author}{\bibfnamefont{R.}~\bibnamefont{Moessner}}, \bibnamefont{and}
	\bibinfo{author}{\bibfnamefont{S.~L.} \bibnamefont{Sondhi}},
	\bibinfo{journal}{Physical review letters} \textbf{\bibinfo{volume}{116}},
	\bibinfo{pages}{250401} (\bibinfo{year}{2016}).
	
	\bibitem[{\citenamefont{Else et~al.}(2016)\citenamefont{Else, Bauer, and
			Nayak}}]{else2016floquet}
	\bibinfo{author}{\bibfnamefont{D.~V.} \bibnamefont{Else}},
	\bibinfo{author}{\bibfnamefont{B.}~\bibnamefont{Bauer}}, \bibnamefont{and}
	\bibinfo{author}{\bibfnamefont{C.}~\bibnamefont{Nayak}},
	\bibinfo{journal}{Physical review letters} \textbf{\bibinfo{volume}{117}},
	\bibinfo{pages}{090402} (\bibinfo{year}{2016}).
	
	\bibitem[{\citenamefont{Yao et~al.}(2017)\citenamefont{Yao, Potter, Potirniche,
			and Vishwanath}}]{yao2017discrete}
	\bibinfo{author}{\bibfnamefont{N.~Y.} \bibnamefont{Yao}},
	\bibinfo{author}{\bibfnamefont{A.~C.} \bibnamefont{Potter}},
	\bibinfo{author}{\bibfnamefont{I.-D.} \bibnamefont{Potirniche}},
	\bibnamefont{and}
	\bibinfo{author}{\bibfnamefont{A.}~\bibnamefont{Vishwanath}},
	\bibinfo{journal}{Physical review letters} \textbf{\bibinfo{volume}{118}},
	\bibinfo{pages}{030401} (\bibinfo{year}{2017}).
	
	\bibitem[{\citenamefont{Martin et~al.}(2017)\citenamefont{Martin, Refael, and
			Halperin}}]{Martin2017}
	\bibinfo{author}{\bibfnamefont{I.}~\bibnamefont{Martin}},
	\bibinfo{author}{\bibfnamefont{G.}~\bibnamefont{Refael}}, \bibnamefont{and}
	\bibinfo{author}{\bibfnamefont{B.}~\bibnamefont{Halperin}},
	\bibinfo{journal}{Phys. Rev. X} \textbf{\bibinfo{volume}{7}},
	\bibinfo{pages}{041008} (\bibinfo{year}{2017}).
	
	\bibitem[{\citenamefont{Huang et~al.}(2018)\citenamefont{Huang, Wu, and
			Liu}}]{Huang-Liu:18}
	\bibinfo{author}{\bibfnamefont{B.}~\bibnamefont{Huang}},
	\bibinfo{author}{\bibfnamefont{Y.-H.} \bibnamefont{Wu}}, \bibnamefont{and}
	\bibinfo{author}{\bibfnamefont{W.~V.} \bibnamefont{Liu}},
	\bibinfo{journal}{Phys. Rev. Lett.} \textbf{\bibinfo{volume}{120}},
	\bibinfo{pages}{110603} (\bibinfo{year}{2018}).
	
	\bibitem[{\citenamefont{{Chen} and {Cai}}(2025)}]{Chen2025}
	\bibinfo{author}{\bibfnamefont{Z.}~\bibnamefont{{Chen}}} \bibnamefont{and}
	\bibinfo{author}{\bibfnamefont{Z.}~\bibnamefont{{Cai}}},
	\bibinfo{journal}{arXiv e-prints} \bibinfo{eid}{arXiv:2503.09437}
	(\bibinfo{year}{2025}), \eprint{2503.09437}.
	
	\bibitem[{\citenamefont{Fu et~al.}(2024)\citenamefont{Fu, Moessner, Zhao, and
			Bukov}}]{Fu2024}
	\bibinfo{author}{\bibfnamefont{Z.}~\bibnamefont{Fu}},
	\bibinfo{author}{\bibfnamefont{R.}~\bibnamefont{Moessner}},
	\bibinfo{author}{\bibfnamefont{H.}~\bibnamefont{Zhao}}, \bibnamefont{and}
	\bibinfo{author}{\bibfnamefont{M.}~\bibnamefont{Bukov}},
	\bibinfo{journal}{Phys. Rev. X} \textbf{\bibinfo{volume}{14}},
	\bibinfo{pages}{041070} (\bibinfo{year}{2024}).
	
	\bibitem[{\citenamefont{Abanin et~al.}(2015)\citenamefont{Abanin, De~Roeck, and
			Huveneers}}]{ExponentiallyDima}
	\bibinfo{author}{\bibfnamefont{D.~A.} \bibnamefont{Abanin}},
	\bibinfo{author}{\bibfnamefont{W.}~\bibnamefont{De~Roeck}}, \bibnamefont{and}
	\bibinfo{author}{\bibfnamefont{F.~m.~c.} \bibnamefont{Huveneers}},
	\bibinfo{journal}{Phys. Rev. Lett.} \textbf{\bibinfo{volume}{115}},
	\bibinfo{pages}{256803} (\bibinfo{year}{2015}).
	
	\bibitem[{\citenamefont{Mori et~al.}(2016)\citenamefont{Mori, Kuwahara, and
			Saito}}]{Rigorous2016Mori}
	\bibinfo{author}{\bibfnamefont{T.}~\bibnamefont{Mori}},
	\bibinfo{author}{\bibfnamefont{T.}~\bibnamefont{Kuwahara}}, \bibnamefont{and}
	\bibinfo{author}{\bibfnamefont{K.}~\bibnamefont{Saito}},
	\bibinfo{journal}{Phys. Rev. Lett.} \textbf{\bibinfo{volume}{116}},
	\bibinfo{pages}{120401} (\bibinfo{year}{2016}).
	
	\bibitem[{\citenamefont{C.Castelnovo et~al.}(2008)\citenamefont{C.Castelnovo,
			R.Moessner, and Sondhi}}]{Castelnovo2008}
	\bibinfo{author}{\bibnamefont{C.Castelnovo}},
	\bibinfo{author}{\bibnamefont{R.Moessner}}, \bibnamefont{and}
	\bibinfo{author}{\bibfnamefont{S.}~\bibnamefont{Sondhi}},
	\bibinfo{journal}{Nature} \textbf{\bibinfo{volume}{451}}, \bibinfo{pages}{42}
	(\bibinfo{year}{2008}).
	
	\bibitem[{\citenamefont{Mori}(2022)}]{Takashi2022Heating}
	\bibinfo{author}{\bibfnamefont{T.}~\bibnamefont{Mori}}, \bibinfo{journal}{Phys.
		Rev. Lett.} \textbf{\bibinfo{volume}{128}}, \bibinfo{pages}{050604}
	(\bibinfo{year}{2022}).
	
	\bibitem[{\citenamefont{Howell et~al.}(2019)\citenamefont{Howell, Weinberg,
			Sels, Polkovnikov, and Bukov}}]{AsymptoticMarin2019}
	\bibinfo{author}{\bibfnamefont{O.}~\bibnamefont{Howell}},
	\bibinfo{author}{\bibfnamefont{P.}~\bibnamefont{Weinberg}},
	\bibinfo{author}{\bibfnamefont{D.}~\bibnamefont{Sels}},
	\bibinfo{author}{\bibfnamefont{A.}~\bibnamefont{Polkovnikov}},
	\bibnamefont{and} \bibinfo{author}{\bibfnamefont{M.}~\bibnamefont{Bukov}},
	\bibinfo{journal}{Phys. Rev. Lett.} \textbf{\bibinfo{volume}{122}},
	\bibinfo{pages}{010602} (\bibinfo{year}{2019}).
	
	\bibitem[{Sup()}]{Supplementary}
	\bibinfo{howpublished}{See the supplementary for some details of the numerical
		simulation, including a derivation of the difference equation to solve the
		stroboscopic spin dynamics and the initial state preparation. We also justify
		the relation between initial state noise strength and the effective
		temperature and derive the high-order terms of the Floquet Hamiltonian in the
		magnus expansion. Finally, we provide an example of the slowly driven case,
		where the system is quickly heated up without a prethermal plateau and
		emergent dynamical frustration.}
	
	\bibitem[{\citenamefont{Ebadi et~al.}(2022)\citenamefont{Ebadi, Keesling, Cain,
			Wang, Levine, Bluvstein, Semeghini, Omran, Liu, Samajdar et~al.}}]{Ebadi2022}
	\bibinfo{author}{\bibfnamefont{S.}~\bibnamefont{Ebadi}},
	\bibinfo{author}{\bibfnamefont{A.}~\bibnamefont{Keesling}},
	\bibinfo{author}{\bibfnamefont{M.}~\bibnamefont{Cain}},
	\bibinfo{author}{\bibfnamefont{T.~T.} \bibnamefont{Wang}},
	\bibinfo{author}{\bibfnamefont{H.}~\bibnamefont{Levine}},
	\bibinfo{author}{\bibfnamefont{D.}~\bibnamefont{Bluvstein}},
	\bibinfo{author}{\bibfnamefont{G.}~\bibnamefont{Semeghini}},
	\bibinfo{author}{\bibfnamefont{A.}~\bibnamefont{Omran}},
	\bibinfo{author}{\bibfnamefont{J.}~\bibnamefont{Liu}},
	\bibinfo{author}{\bibfnamefont{R.}~\bibnamefont{Samajdar}},
	\bibnamefont{et~al.}, \bibinfo{journal}{Science}
	\textbf{\bibinfo{volume}{376}}, \bibinfo{pages}{1290} (\bibinfo{year}{2022}).
	
	\bibitem[{\citenamefont{Bluvstein et~al.}(2023)\citenamefont{Bluvstein, Evered,
			Geim, Li, Zhou, Manovitz, Ebadi, Cain, Kalinowski, Hangleiter
			et~al.}}]{Bluvstein2023}
	\bibinfo{author}{\bibfnamefont{D.}~\bibnamefont{Bluvstein}},
	\bibinfo{author}{\bibfnamefont{S.~J.} \bibnamefont{Evered}},
	\bibinfo{author}{\bibfnamefont{A.~A.} \bibnamefont{Geim}},
	\bibinfo{author}{\bibfnamefont{S.~H.} \bibnamefont{Li}},
	\bibinfo{author}{\bibfnamefont{H.}~\bibnamefont{Zhou}},
	\bibinfo{author}{\bibfnamefont{T.}~\bibnamefont{Manovitz}},
	\bibinfo{author}{\bibfnamefont{S.}~\bibnamefont{Ebadi}},
	\bibinfo{author}{\bibfnamefont{M.}~\bibnamefont{Cain}},
	\bibinfo{author}{\bibfnamefont{M.}~\bibnamefont{Kalinowski}},
	\bibinfo{author}{\bibfnamefont{D.}~\bibnamefont{Hangleiter}},
	\bibnamefont{et~al.}, \bibinfo{journal}{Nature}
	\textbf{\bibinfo{volume}{626}}, \bibinfo{pages}{58} (\bibinfo{year}{2023}).
	
	\bibitem[{\citenamefont{Hallen et~al.}(2022)\citenamefont{Hallen, Grigera,
			Tennant, Castelnovo, and Moessner}}]{Hallen2022}
	\bibinfo{author}{\bibfnamefont{J.~N.} \bibnamefont{Hallen}},
	\bibinfo{author}{\bibfnamefont{S.~A.} \bibnamefont{Grigera}},
	\bibinfo{author}{\bibfnamefont{D.~A.} \bibnamefont{Tennant}},
	\bibinfo{author}{\bibfnamefont{C.}~\bibnamefont{Castelnovo}},
	\bibnamefont{and} \bibinfo{author}{\bibfnamefont{R.}~\bibnamefont{Moessner}},
	\bibinfo{journal}{Science} \textbf{\bibinfo{volume}{378}},
	\bibinfo{pages}{1218} (\bibinfo{year}{2022}).
	
	\bibitem[{\citenamefont{Jaubert and Holdsworth}(2009)}]{Jaubert2009}
	\bibinfo{author}{\bibfnamefont{L.}~\bibnamefont{Jaubert}} \bibnamefont{and}
	\bibinfo{author}{\bibfnamefont{P.}~\bibnamefont{Holdsworth}},
	\bibinfo{journal}{Nat. Phys.} \textbf{\bibinfo{volume}{5}},
	\bibinfo{pages}{258} (\bibinfo{year}{2009}).
	
	\bibitem[{\citenamefont{Castelnovo et~al.}(2010)\citenamefont{Castelnovo,
			Moessner, and Sondhi}}]{Castelnovo2010}
	\bibinfo{author}{\bibfnamefont{C.}~\bibnamefont{Castelnovo}},
	\bibinfo{author}{\bibfnamefont{R.}~\bibnamefont{Moessner}}, \bibnamefont{and}
	\bibinfo{author}{\bibfnamefont{S.~L.} \bibnamefont{Sondhi}},
	\bibinfo{journal}{Phys. Rev. Lett.} \textbf{\bibinfo{volume}{104}},
	\bibinfo{pages}{107201} (\bibinfo{year}{2010}).
	
	\bibitem[{\citenamefont{Bramwell et~al.}(2009)\citenamefont{Bramwell, Giblin,
			Calder, Aldus, Prabhakaran, and Fennell}}]{Bramwell2009}
	\bibinfo{author}{\bibfnamefont{S.~T.} \bibnamefont{Bramwell}},
	\bibinfo{author}{\bibfnamefont{S.}~\bibnamefont{Giblin}},
	\bibinfo{author}{\bibfnamefont{S.}~\bibnamefont{Calder}},
	\bibinfo{author}{\bibfnamefont{R.}~\bibnamefont{Aldus}},
	\bibinfo{author}{\bibfnamefont{D.}~\bibnamefont{Prabhakaran}},
	\bibnamefont{and} \bibinfo{author}{\bibfnamefont{T.}~\bibnamefont{Fennell}},
	\bibinfo{journal}{Nature} \textbf{\bibinfo{volume}{461}},
	\bibinfo{pages}{956} (\bibinfo{year}{2009}).
	
	\bibitem[{\citenamefont{S.Mostame et~al.}(2014)\citenamefont{S.Mostame,
			C.Castelnovo, R.Moessner, and Sondhi}}]{Mostame2014}
	\bibinfo{author}{\bibnamefont{S.Mostame}},
	\bibinfo{author}{\bibnamefont{C.Castelnovo}},
	\bibinfo{author}{\bibnamefont{R.Moessner}}, \bibnamefont{and}
	\bibinfo{author}{\bibfnamefont{S.}~\bibnamefont{Sondhi}},
	\bibinfo{journal}{PNAS} \textbf{\bibinfo{volume}{2}}, \bibinfo{pages}{640}
	(\bibinfo{year}{2014}).
	
	\bibitem[{\citenamefont{Pan et~al.}(2016)\citenamefont{Pan, Laurita, Ross,
			Gaulin, and Armitage}}]{Pan2016}
	\bibinfo{author}{\bibfnamefont{L.}~\bibnamefont{Pan}},
	\bibinfo{author}{\bibfnamefont{N.~J.} \bibnamefont{Laurita}},
	\bibinfo{author}{\bibfnamefont{K.~A.} \bibnamefont{Ross}},
	\bibinfo{author}{\bibfnamefont{B.~D.} \bibnamefont{Gaulin}},
	\bibnamefont{and} \bibinfo{author}{\bibfnamefont{N.~P.}
		\bibnamefont{Armitage}}, \bibinfo{journal}{Nat. Phys}
	\textbf{\bibinfo{volume}{12}}, \bibinfo{pages}{361} (\bibinfo{year}{2016}).
	
	\bibitem[{\citenamefont{Lubelski et~al.}(2008)\citenamefont{Lubelski, Sokolov,
			and Klafter}}]{Igor2008}
	\bibinfo{author}{\bibfnamefont{A.}~\bibnamefont{Lubelski}},
	\bibinfo{author}{\bibfnamefont{I.~M.} \bibnamefont{Sokolov}},
	\bibnamefont{and} \bibinfo{author}{\bibfnamefont{J.}~\bibnamefont{Klafter}},
	\bibinfo{journal}{Phys. Rev. Lett.} \textbf{\bibinfo{volume}{100}},
	\bibinfo{pages}{250602} (\bibinfo{year}{2008}).
	
	\bibitem[{\citenamefont{Bak et~al.}(1987)\citenamefont{Bak, Tang, and
			Wiesenfeld}}]{Bak1987}
	\bibinfo{author}{\bibfnamefont{P.}~\bibnamefont{Bak}},
	\bibinfo{author}{\bibfnamefont{C.}~\bibnamefont{Tang}}, \bibnamefont{and}
	\bibinfo{author}{\bibfnamefont{K.}~\bibnamefont{Wiesenfeld}},
	\bibinfo{journal}{Phys. Rev. Lett.} \textbf{\bibinfo{volume}{59}},
	\bibinfo{pages}{381} (\bibinfo{year}{1987}).
	
	\bibitem[{\citenamefont{Homeier et~al.}(2025)\citenamefont{Homeier, Pizzi,
			Zhao, Halimeh, Grusdt, and Rey}}]{homeier2025prethermal}
	\bibinfo{author}{\bibfnamefont{L.}~\bibnamefont{Homeier}},
	\bibinfo{author}{\bibfnamefont{A.}~\bibnamefont{Pizzi}},
	\bibinfo{author}{\bibfnamefont{H.}~\bibnamefont{Zhao}},
	\bibinfo{author}{\bibfnamefont{J.~C.} \bibnamefont{Halimeh}},
	\bibinfo{author}{\bibfnamefont{F.}~\bibnamefont{Grusdt}}, \bibnamefont{and}
	\bibinfo{author}{\bibfnamefont{A.~M.} \bibnamefont{Rey}},
	\bibinfo{journal}{arXiv preprint arXiv:2510.12800}  (\bibinfo{year}{2025}).
	
\end{thebibliography}

\end{document}